\documentstyle{elsart}

\begin{document}
\begin{frontmatter}
\title{Characterising a Si(Li) detector element for the SIXA X-ray
 spectrometer}

\author{T.~Tikkanen}
\address{Observatory and Astrophysics Laboratory,
 P.O. Box 14 (T\"ahti\-tornin\-m\"aki),
 FIN-00014 University of Helsinki,
 Finland}
\author{S.~Kraft},
\author{F.~Scholze},
\author{R.~Thornagel},
\author{G.~Ulm}
\address{Physikalisch-Technische Bundesanstalt,
 Abbestr.\ 2--12, D-10587 Berlin, Germany}

\begin{abstract}
The detection efficiency and response function of a Si(Li) detector
 element for the SIXA spectrometer have been determined in the 500~eV to
 5~keV energy range using synchrotron radiation emitted at a bending
 magnet of the electron storage ring BESSY, which is a primary radiation
 standard. The agreement between the
 measured spectrum and the model calculation is better than 2\%.

\end{abstract}

\end{frontmatter}

\section{Introduction}

The SIXA (Silicon X-Ray Array) spectrometer \cite{Vilhu} is a focal
 plane instrument of the SODART X-ray telescope on board the Russian
 Spectrum-X-Gamma satellite scheduled for launch in 1998. SIXA is a
 closely packed array of 19 discrete Si(Li) detector elements which
 collect X-rays in the energy range between 500~eV and 20~keV with an
 energy resolution of 200~eV at 6~keV\@. The detector crystals are kept
 at a temperature of about 120~K by a passive cooling system.

Although simple calibration methods involving a set of discrete X-ray
 lines from radioactive sources may be adequate for many applications
 of Si(Li) detectors, this is not the case in X-ray astronomy where one
 often wants to resolve emission or absorption line features superposed
 on a continuum. Fine structures in the instrumental response associated
 with absorption edges of the constituent elements of the instrument can
 mimic such line features, and these spurious lines may coincide with
 real lines from astronomical sources. Modern instruments combining high
 throughput X-ray optics with detectors of high resolving power have
 proved to be able to resolve such fine effects: for example, the
 spectrum of the Crab Nebula (which normally has no features) had a hump
 at the L~absorption edge of xenon when observed by the xenon-filled
 GSPC's on board the Japanese Tenma satellite \cite{Koyama} and line
 features near the K~edges of silicon, aluminium and oxygen when
 observed by NASA's Broad Band X-Ray Telescope which contained these
 elements in its segmented Si(Li) detector and entrance window
 \cite{Weaver,Serlemitsos}.

Synchrotron radiation (SR) is needed to resolve X-ray absorption fine
 structure (XAFS) or to obtain accurate characterisation in the photon
 energy range below 5~keV where other appropriate X-ray sources are not
 available. Soft X-ray transmission of the entrance window of SIXA was
 measured using SR \cite{Tikkanen}. XAFS of the detector surface has
 been measured recently and is to be published in a future paper.
 Simulations suggested that the effect of the entrance surface on the
 detection process in Si(Li) detectors, as modelled by Scholze and Ulm
 \cite{Scholze}, is of particular importance for SIXA in the range below
 3~keV\@. In this paper we report the characterisation of a SIXA
 detector element carried out following their procedure, which resulted
 in the experimental determination of the detector response and the
 detection efficiency. Another goal of this experiment was to
 investigate a temperature dependent low-energy tailing phenomenon
 reported by the manufacturer (Metorex International Oy, Espoo,
 Finland).

\section{Detector model}

\subsection{Detector design}

The detector elements fabricated for SIXA are top-hat Si(Li) detectors
 which are 3.5~mm thick and have an active diameter of 9.2~mm (see
 Fig.~\ref{fig:sili}). X-rays enter through a contact layer of Au/Pd
 alloy whose nominal thickness is 30~nm with a mass composition of
 60\% Au and 40\% Pd. The anode side of the crystal has a smaller
 diameter to reduce the readout capacitance, thus yielding a better
 energy resolution. The diameter is greater on the cathode side where
 the drifted region is encircled by uncompensated {\em p}-type silicon
 which has been left there to facilitate the handling of the crystal.
 The edge is coated with polyimide to provide passivation.

\subsection{X-ray response}

The X-ray response of the SIXA elements was modelled according to
 ref.~\cite{Scholze} where a detailed description of the detector model
 is given, so that here only a summary of the basic concepts of the
 model is presented. The main feature of the model is that no inactive
 layer of silicon is assumed; instead, the "window effect" is explained
 by a strong expansion of the charge cloud before thermalisation and
 consequent escape of electrons into the contact material.

A photon impinging on the detector can either be transmitted or absorbed
 in the detector crystal, the contact material or possible contamination
 layers such as carbon and oxygen. Absorption in the contact layer
 produces photo- and Auger electrons and fluorescence photons which can
 be lost without being detected or, with a calculable probability, enter
 the crystal. Every photon absorbed in the crystal and every electron
 entering the crystal produces a cloud of electron--hole pairs in the
 Si(Li) crystal. The charge carriers are thermalised and start to drift
 in the electric field set up in the detector, which produces a charge
 signal on the electrodes. The detection efficiency is defined as a
 total detection efficiency $\epsilon(E)$ which includes all pulses
 produced by photons absorbed in the detector.

The total detection efficiency $\epsilon(E)$ of a SIXA crystal is equal
 to the absorptance of the active region times the transmittance of the
 Au/Pd contact and the contamination layers plus the probability
 $P_{\mathrm{Au/Pd}}(E)$ that an incident photon absorbed in the
 Au/Pd alloy will produce a pulse:
 \begin{equation} \label{eq:tot}
 \epsilon(E) = (1 - \tau_{\mathrm{Si}}) \tau_{\mathrm{Au/Pd}}
 \tau_{\mathrm{C}} \tau_{\mathrm{ice}} + P_{\mathrm{Au/Pd}} .
 \end{equation}
 As indicated above, the latter contribution originates from absorption
 events where a photo- or Auger electron or a fluorescence photon is
 generated in the contact layer and emitted into the crystal. Unlike the
 full-peak detection efficiency, which includes only the Gaussian part
 of the peak, $\epsilon(E)$ is independent of the attenuation
 coefficient of silicon at low energies.

The pulse height distribution $C(E)$ measured with the detector exposed
 to the spectral photon flux $\Phi_E(E)$ is given by
\begin{equation} \label{eq:c}
 C(E) = \int R(E^\prime,E) \epsilon(E^\prime) \Phi_E(E^\prime)
 \d E^\prime .
\end{equation}
Knowledge of the normalised response function $R(E^\prime,E)$ and the
 photon flux $\Phi_E(E)$ facilitates the determination of $\epsilon(E)$
 from the measured pulse height distribution $C(E)$.

\subsection{Simulation}

The influence of the detector model on the analysis of astronomical data
 acquired by SIXA was investigated by simulating an observation of a
 typical astronomical target by SIXA, following the procedure described
 in ref.~\cite{Tikkanen} where it was applied to XAFS in the entrance
 window. The same power-law spectrum modified by interstellar absorption
 (simulating the Crab Nebula) was folded through the combined
 instrumental response of SIXA and SODART to compute the simulated data.
 The data were then modelled in a traditional way assuming a dead layer
 of silicon to explain the window effect; the response function in this
 model consisted simply of two Gaussian peaks, the full-energy peak and
 the escape peak. The result is presented in Fig.~\ref{fig:crab}. The
 dead layer thickness was 160~nm which would have been the result of a
 measurement with a \nuc{55}Fe source.

Although the dead layer model can reproduce the data very well at higher
 energies, it introduces spurious structures at the K~edge of silicon
 and at the M~edges of gold, and below 1~keV the models are completely
 in disaccord. The result indicates clearly that if full scientific
 return at low energies is required, the detector response has to be
 accurately determined using an appropriate model with parameters
 abstracted from SR calibration data.

\section{Experiment}

The SR measurements were performed at the PTB radiometry laboratory at
 the electron storage ring BESSY\@. The SX700 plane grating
 monochromator of the PTB radiometry laboratory \cite{Scholze2} was
 utilized to measure the response function in the 0.6~keV to 1.5~keV
 photon energy range and the double crystal monochromator KMC of BESSY
 \cite{Feldhaus} was used in the 1.8~keV to 5.9~keV energy range. The
 measurements with undispersed SR were carried out at a specially
 designed beamline of the PTB radiometry laboratory. The photon flux of
 undispersed SR emitted at a bending magnet of the electron storage ring
 BESSY is calculable with an uncertainty well below 0.5\% in the desired
 energy range \cite{Rabus,Schwinger,Arnold}. The accuracy of the
 calibration is mainly limited by the ability to extract the thicknesses
 of the contact and contamination layers. Because of the moderate
 resolution of the detector, fine structures cannot be recovered.

The SIXA flight assembly was being assembled during the measurement
 shift at BESSY and therefore the characterisation had to be done using
 one of three available crystals which were left over after the
 selection of the best crystals for the flight model. The three crystals
 were studied at Metorex using an electron microscope in order to select
 the best representative of a typical flight model crystal with respect
 to the temperature dependent tailing effect. Many crystals appeared to
 have a critical temperature (which usually fell near the expected
 in-orbit operation temperature) where the low-energy tail started to
 grow rapidly with temperature. One of the three crystals did not
 exhibit this effect even at 170~K, while another one was found to have
 suffered from shelf storage. The third one was suitable: the tail
 between the main peak and the escape peak became about two times higher
 when the temperature was raised from 125~K to 130~K\@. This crystal was
 chosen to be characterised with SR and to serve as transfer standard
 detector for calibration of the other SIXA crystals.

\section{Measurement and modelling of the response function}

\subsection{Homogeneity}

The detector homogeneity was tested by positioning the beam at five
 different locations on the crystal (cf.\ Fig.~\ref{fig:sili}). At the
 position '0~mm' the beam was located at the edge of the active region
 so that half of the total intensity was detected. As can be seen in
 Fig.~\ref{fig:homog} the response function was very similar at the
 three central positions, while the tail is much higher at the '0~mm'
 position. At the '8.2~mm' position near the opposite edge, a slight
 increase of the tailing is observable. The temperature was 124~K and
 the beam width about 1~mm. The temperature dependence was studied with
 the beam positioned at the centre. No change in the tail structure was
 seen up to the temperature of 132~K, although noise and the FWHM
 increased as expected (see Fig.~\ref{fig:temp}).

These results suggest that the previously observed growth of the tailing
 between 125 and 130~K, which was measured using an isotropic source and
 an aperture diameter of 9.2~mm, occurs only at the periphery of the
 detector. Top-hat detectors typically suffer from tailing in peripheral
 regions where the signal electrons drift towards the side surface
 rather than the anode because an {\em n}-type channel is formed on the
 surface \cite{Jaklevic}. Despite their polyimide passivation, SIXA
 crystals are obviously subject to this effect as well. Leakage current
 is generated on the same surface and the leakage currents were found to
 be as high in polyimide coated crystals as in uncoated crystals
 \cite{Jantunen}. The sensitivity of the surface potential to ambient
 conditions provides an obvious explanation for the observed temperature
 dependence of the tailing. This is illustrated in
 Fig.~\ref{fig:equipot} which depicts the detector at two different
 temperatures. The potential profiles were computed by the
 two-dimensional modelling program SCORPIO \cite{Grahn} with different
 effective doping densities of the surface channel. Photon absorptions
 at the periphery yield defective pulses because electron clouds
 generated at greater radial distances drift to the side surface where
 electrons can be trapped. Tailing is more pronounced in the situation
 of the lower plot where the effective doping is higher, corresponding
 to a higher temperature. The result of the homogeneity test can be
 understood with the upper plot. The '0~mm' position was at a radial
 distance of about 4.6~mm. Electron paths starting around this distance
 end at the side surface, thus explaining the tail. On the other hand,
 electrons from around 3.6~mm head towards the back surface outside the
 anode, producing pulses with almost full energy which cause the small
 broadening of the left side of the peak at the '8.2~mm' position. The
 tail feature further from the peak is caused by the part of the beam
 extending closer to the edge.

\subsection{Response function}

The measured response functions were fitted by using the HYPERMET
 function in the form of \cite{Scholze} including a Gaussian peak, an
 escape peak, an exponential tail and a flat shelf. The total function
 includes ten energy dependent parameters. The shelf contribution is
 attributed to the escape probabilities of primary electrons to and from
 the contact layer (Au/Pd alloy). The calculated and fitted shelf
 contribution can be seen in Fig.~\ref{fig:shelf}; the calculation
 includes the most probable Auger and photoelectron energies. The short
 tail is caused by the escape of hot electrons into the contact layer.
 Fig.~\ref{fig:tail} shows this contribution as a function of energy and
 a fit with $R=210$~nm, where $R$ is the radius of the spherical
 electron cloud. It can be seen that below 1.8~keV the difference to the
 fitted tail contribution increases with decreasing energy. The
 measurement in this energy region was taken at the plane grating
 monochromator, where more stray light exists at energies above 1~keV
 resulting in higher tailing contributions. At lower energies the peak
 cannot properly be extracted from the noise leading to fits with lower
 tail contributions. These difficulties demonstrate the necessity of a
 theoretical model which allows an extrapolation of the fit parameters
 to lower energies.

Using these parameters the response function in the 500~eV to 4~keV
 range can be constructed for the central region. Fig.~\ref{fig:resp}
 shows a comparison of some typical measured distributions and the
 corresponding theoretical curves described by the response function.
 The inhomogeneity of the response should be taken into account,
 because the aperture diameter in the flight model array will be about
 9~mm and thus the excess tailing can affect a great part of the active
 area. Excluding the peripheral region by additional collimation is to
 be avoided as it would diminish the effective area. At 124~K about
 10\% of the area seems to be affected in the present case, but
 Fig.~\ref{fig:equipot} suggests that the affected area can be 2 or 3
 times larger at depths of 1--2~mm where more energetic photons would
 be absorbed. The affected region can be expected to grow with the
 temperature, and the region will have a different size for each Si(Li)
 crystal.

\section{Calibration with undispersed synchrotron radiation}

For the measurements with undispersed SR the number of stored electrons
 was decreased to either 5 or 2 electrons, yielding respectively about
 3600 and 1500 photons
 per second striking the detector. The flux through an aperture with an
 area of 27.8(2)~mm$^2$ at a distance of 15783(3)~mm from the source
 point was calculated from the known electron storage ring parameters
 \cite{Rabus} in the energy range 100~eV to 5~keV\@. The measured
 spectra were compared to the model calculations of Eq.~(\ref{eq:c}).
 The determined response function of the central area is valid for this
 measurement because the aperture was small enough and the coldfinger
 temperature was 125~K\@.

A proper comparison of the measured spectra and the predicted spectra
 requires an energy calibration of the multichannel analyser with an
 uncertainty of about 0.1\%. For the energy calibration the line
 position at 900~eV was determined by the Gaussian peak position of the
 best fit of the response function. The SX700 energy scale is more
 accurate than 0.5~eV at this point. At 6.4~keV the Fe~K$_\alpha$
 emission line(s) of a \nuc{55}{Co} source was used for the calibration.
 A linear gain was assumed.

A considerable pile-up contribution in the spectra measured with 2 and 5
 electrons in the electron storage ring was observed. The pile-up rate 
 or coincidence probability of two pulses with the count rates $N(E_1)$
 and $N(E_2)$ occuring in the interval $T_{\mathrm R}$ is equal to the
 product \cite{Knoll},
\begin{equation}
N_{\mathrm{P}}(E_1 + E_2) = N(E_1) N(E_2) T_{\mathrm R} ,
\end{equation}
if $T_{\mathrm R} N(E) \ll 1$. The resolving time $T_{\mathrm R}$ is in
 first approximation a constant. The pile-up contribution
 $N_{\mathrm{P}}(E)$ of the whole spectrum can be extracted via the
 calculation of the auto-correlation function of the calculated spectrum
 \cite{Tenney,Datlowe}:
\begin{equation}
\label{pile-up}
N_{\mathrm{P}}(E) \sim
  \int_0^\infty C(E^\prime) C(E-E^\prime) \d E^\prime .
\end{equation}

Two photons impinging within the time interval $T_{\mathrm{R}}$ cannot
 be resolved by the electronics and appear as one pile-up pulse. From
 the difference of the spectra taken with 2 and 5 electrons in the
 electron storage ring $T_{\mathrm R}$ can be determined in the way that
 with the appropriate $T_{\mathrm R}$ both spectra coincide after a
 pile-up correction. $T_{\mathrm R}$ can be used as a proportionality
 constant for Eq.~(\ref{pile-up}). The doubled number of calculated
 pile-up pulses $N_{\mathrm{P}}$ have to be subtracted afterwards, so
 that the pile-up corrected spectrum $C^\prime(E)$ is
\begin{equation}
C^\prime(E) =
C(E) \left( 1 - 2 \frac{N_{\mathrm{P}}}{N} \right) + N_{\mathrm{P}}(E) .
\label{pileup}
\end{equation}
Applying this formula to the present spectra results in a constant
 relative deviation of 3\% for the 2 electron spectrum and of about 10\%
 for the 5 electron spectrum. An explanation of the result might be that
 in the real detection process the deadtime is overestimated. An
 influence of low energy pulses which are part of the calculation but
 cannot be seen by the electronics might also be possible. The deadtime
 of the measurements was 11\% and 23\% with 2 and 5 electrons
 respectively. To account for a realistic pile-up distribution, the
 numerical calculated pile-up pulses are not only distributed in the sum
 energy corresponding channel, but also in all channels between both
 contributing pulses. This has been taken into account for the present
 calculations. In order to obtain two coincident spectra the factor 2 in
 Eq.~(\ref{pileup}) has to be replaced by $\sqrt{2}$.
 Fig.~\ref{fig:pileup} shows the comparison of the measured and
 calculated spectra. A resolving time of 25~$\mu$s has been used in
 order to fit the measurements.

If the pile-up rejection works, the deadtime and the pile-up influence
 is neglectable for a typical photon flux of a few hundred per second.
 The excellent agreement within the statistical uncertainty in the
 energy range 500~eV to 4~keV for both the low count rate spectrum and
 the high count rate spectrum confirms the correctness of the
 calculation in this particular case. The lowest deviation is found with
 thicknesses of 20.1(3)~nm and 20.1(3)~nm for Au and Pd, respectively,
 and an ice layer of 16(3)~nm. This is equivalent to a contact layer
 thickness of 40.7~nm and nearly consistent with the mass ratio of 60:40
 of the elements Au and Pd. A possible carbon layer can be neglected.
 The detection efficiency calculated with the determined parameters from
 Eq.~(\ref{eq:tot}) as well as the full-peak efficiency are shown in
 Fig.~\ref{fig:de}. The uncertainty reflects the thickness determination
 of the Au, Pd and ice layers and the shelf. The accuracy of the total
 photon flux is limited by the knowledge of the detector aperture size.
 It should be mentioned that the uncertainties are determined on the
 basis of results obtained with different models of pile-up calculations
 and would have been lower without any pile-up effect.

\section{Conclusion}

With the aid of dispersed and undispersed SR, the response function and
 the detection efficiency in the central region of a Si(Li) crystal for
 the SIXA spectrometer in the soft X-ray range have been determined. The
 agreement between the calculated and the measured spectra within 2\% is
 a further confirmation of the correctness of the physical detector
 model. The detection efficiency has been determined with an uncertainty
 below 1.5\% above 1~keV\@. Although excellent results have been
 obtained after a pile-up correction, a repetition of the measurement
 with undispersed SR is recommended, because the characterised detector
 element will be taken as a transfer standard for the calibration of the
 SIXA flight assembly.

The previously observed temperature dependence was found to arise from
 the inhomogeneity of the response near the edge of the cylindrical
 crystal. This effect is presumed to cover a large part of the detector
 area and requires therefore further study.

\begin{ack}
We thank M. Jantunen of Metorex for information about the crystals
 and the test results.
\end{ack}

\newpage

\begin{figure}
\caption{Cross section of a Si(Li) detector element of the SIXA array.
 The compensated region is denoted by hatching. Beam positions of the
 homogeneity test (Fig.~\protect\ref{fig:homog}) are indicated.}
\label{fig:sili}
\end{figure}

\begin{figure}
\caption{Illustrating the significance of detector characterisation with
 simulated observations of the Crab Nebula by SIXA/SODART: simulated
 data from a 5000~s observation (points), obtained by folding an
 absorbed power-law model spectrum through the predicted instrumental
 response matrix, are compared to a model (curve) computed using a
 simpler detector model with a dead layer. The curve in the lower panel
 represents residuals with infinite observation time.}
\label{fig:crab}
\end{figure}

\begin{figure}
\caption{Variation of the detector response function across the crystal
 surface. The spectra obtained at the positions 2.3~mm, 4.6~mm and
 6.9~mm almost coincide.}
\label{fig:homog}
\end{figure}

\begin{figure}
\caption{The measured response function at 2.68~keV with coldfinger
 temperatures of 128~K (solid curve) and 99~K (dotted curve) normalised
 to the number of counts above 1~keV\@.}
\label{fig:temp}
\end{figure}

\begin{figure}
\caption{Equipotential curves (eV) and drift paths of electrons in the
 detector at different temperatures (upper plot depicts a lower
 temperature).}
\label{fig:equipot}
\end{figure}

\begin{figure}
\caption{Contribution of the shelf, measured (points) and calculated
 (curve).}
\label{fig:shelf}
\end{figure}

\begin{figure}
\caption{Contribution of the short tail as fitted by the measured
 response functions compared to the best-fit model calculation with
 $R=210$~nm. Usage of a theoretical model overcomes the larger
 uncertainties of the measurement at low energies like here below
 1.8~keV (see text).}
\label{fig:tail}
\end{figure}

\begin{figure}
\caption{Typical response functions at low energy and near and far above
 the Si K absorption edge compared to the model functions.}
\label{fig:resp}
\end{figure}

\begin{figure}
\caption{Upper figure: Comparison of the measured pulse height
 distributions taken with an electron current of 2 and 5 electrons in
 the storage ring and the calculations including the pile-up
 contribution. C(E) is the number of counts or photons per stored
 electron and eV\@. The dashed curve indicates the calculated spectrum
 without pile-up. Lower figure: Relative difference to the calculations
 for the measurements with 5 (crosses) and 2 (diamonds) electrons stored
 in the electron storage ring. The statistical uncertainty is indicated
 by the solid lines.}
\label{fig:pileup}
\end{figure}

\begin{figure}
\caption{Upper part: Determined detection efficiency (DE) of the Si(Li)
 detector according to the transmittance of 20.1(3)~nm Au, 20.1(3)~nm Pd
 and 16(3)~nm ice and the calculated shelf contribution. The dashed
 curve denotes the full-peak efficiency. Lower part: Corresponding total
 uncertainty including the transmittance uncertainty (dashed line), the
 shelf uncertainty (dashed-dotted line) and the uncertainty of the
 aperture size.}
\label{fig:de}
\end{figure}


\begin{thebibliography}{99}

\bibitem{Vilhu} O. Vilhu, J. Huovelin, T. Tikkanen, P. Hakala, P. Muhli,
 V.J. K\"am\"ar\"ainen, H. Sipil\"a, I. Taylor, J. Pohjonen,
 H. P\"aivike, J. Toivanen, R. Sunyaev, A. Kuznetsov and A. Abrosimov,
 Proc.\ SPIE 2279 (1994) 532.

\bibitem{Koyama} K. Koyama, T. Ikegami, H. Inoue, N. Kawai,
 K. Makishima, M. Matsuoka, K. Mitsuda, T. Murakami, Y. Ogawara,
 T. Ohashi, K. Suzuki, Y. Tanaka, I. Waki and E.E. Fenimore,
 Publ.\ Astron.\ Soc.\ Japan 36 (1984) 659.

\bibitem{Weaver} K.A. Weaver, Legacy - The Journal of the HEASARC,
 Nr.~5 (1994) 12.

\bibitem{Serlemitsos} P.J. Serlemitsos, F.E. Marshall, R. Petre,
 K. Jahoda, E.A. Boldt, S.S. Holt, R. Mushotzky, J. Swank,
 A. Szymkowiak, R. Kelley and M. Loewenstein,
 in Frontiers of X-Ray Astronomy.
 28\raisebox{1ex}{\scriptsize th} Yamada meeting,
 eds.\ Y. Tanaka and K. Koyama,
 Tokyo: Universal Academy Press, 1991, p.~221.

\bibitem{Tikkanen} T. Tikkanen and J. Huovelin,
 Nucl.\ Instr.\ and Meth.\ A 379 (1996) 130.

\bibitem{Scholze} F. Scholze and G. Ulm,
 Nucl.\ Instr.\ and Meth.\ A 339 (1994) 49.

\bibitem{Scholze2} F. Scholze, M. Krumrey, P. M\"uller and D. Fuchs,
 Rev.\ Sci.\ Instrum.\ 65 (1994) 3229.

\bibitem{Feldhaus} J. Feldhaus, F. Sch\"afers and W. Peatman,
 Proc.\ SPIE 733 (1986) 242.

\bibitem{Rabus} H. Rabus, F. Scholze, R. Thornagel and G. Ulm,
 Nucl.\ Instr.\ and Meth.\ A 377 (1996) 209.

\bibitem{Schwinger} J. Schwinger, Phys.\ Rev.\ 75 (1948) 1912.

\bibitem{Arnold} D. Arnold and G. Ulm,
 Rev.\ Sci.\ Instrum.\ 60 (1989) 2287.

\bibitem{Jaklevic} J.M. Jaklevic and F.S. Goulding,
 IEEE Trans.\ Nucl.\ Sci.\ 19 (1972) 384.

\bibitem{Jantunen} M. Jantunen and S.A. Audet,
 Nucl.\ Instr.\ and Meth.\ A 353 (1994) 89.

\bibitem{Grahn} K.J. Grahn, Acta Polytechnica Scandinavica, El.\ 76
 (1993) 1.

\bibitem{Knoll} G.F. Knoll, Radiation detection and measurement,
 2\raisebox{1ex}{\scriptsize nd} ed., Wiley, New York 1989, p.~304.

\bibitem{Tenney} F.H. Tenney, Nucl.\ Instr.\ and Meth.\ 219 (1984) 165.

\bibitem{Datlowe} D.W. Datlowe, Nucl.\ Instr.\ and Meth.\ 145 (1977)
 365.

\end{thebibliography}
\end{document}